\documentclass[reqno,11pt]{amsart}
\usepackage{hyperref}
\usepackage[mathscr]{eucal}
\numberwithin{equation}{section}

\title[The Dirac Equation and the Normalization in a Closed FRW Universe]{The Dirac Equation and the Normalization of its Solutions in a Closed Friedmann-Robertson-Walker Universe}

\author[F.\ Finster]{Felix Finster}
\thanks{First author supported in part by the Deutsche Forschungsgemeinschaft.}
\address{NWF I - Mathematik \\ Universit\"at Regensburg \\ D-93040 Regensburg \\ Germany}
\email{Felix.Finster@mathematik.uni-regensburg.de}

\author[M.\ Reintjes]{Moritz Reintjes \\ \\ January 2009}
\address{Mathematics Department \\ University of California, Davis \\ Davis, CA 95616, USA}
\email{moritz@math.ucdavis.edu}

\newtheorem{Def}{Definition}[section]
\newtheorem{Thm}[Def]{Theorem}
\newtheorem{Prp}[Def]{Proposition}
\newtheorem{Lemma}[Def]{Lemma}

\newcommand{\Thanks}{\vspace*{.5em} \noindent \thanks}
\newcommand{\beq}{\begin{equation}}
\newcommand{\eeq}{\end{equation}}
\newcommand{\Proof}{\begin{proof}}
\newcommand{\QED}{\end{proof} \noindent}

\newcommand{\bra}{\mbox{$< \!\!$ \nolinebreak}}
\newcommand{\ket}{\mbox{\nolinebreak $>$}}
\newcommand{\C}{\mathbb{C}}
\newcommand{\R}{\mathbb{R}}
\newcommand{\1}{\mbox{\rm 1 \hspace{-1.05 em} 1}}

\newcommand{\N}{\mathbb{N}}
\newcommand{\D}{\mathcal{D}}
\newcommand{\Pdd}{\mbox{$\partial$ \hspace{-1.2 em} $/$}}
\newcommand{\slsh}{\mbox{ \hspace{-1.1 em} $/$}}

\allowdisplaybreaks[1]

\begin{document}

\begin{abstract}
We set up the Dirac equation in a Friedmann-Robertson-Walker geometry and separate
the spatial and time variables. In the case of a closed universe,
the spatial dependence is solved explicitly, giving rise to a discrete set of solutions.
We compute the probability integral and analyze a space-time normalization integral.
This analysis allows us to introduce the fermionic projector in a closed Friedmann-Robertson-Walker geometry and to specify its global normalization as well as its local form.
\end{abstract}
\maketitle

\tableofcontents

\section{Introduction}
General relativity and quantum theory describe physics on different scales.
Whereas the large-scale structure of our universe is governed by Einstein's general theory of relativity,
on the small scale gravitational forces are often negligible,
and instead physical effects are described by quantum theory.
Due to the different scales, in most situations it is unnecessary to combine the two theories
(and in fact, a convincing ``unification'' of general relativity and quantum theory is still unknown).
Nevertheless, there are issues in quantum theory where the large scale structure of space-time
does become relevant, so that general relativity must be taken into account. Such problems are usually
referred to as infrared problems. To give a simple example,
for the {\em{normalization}} of a wave function one must integrate the probability density over
all of space, and thus the analysis of the normalization integral depends sensitively
on the global structure of space.
Moreover, in relativistic quantum theory,
four-dimensional integrals of the wave functions over both space and time appear.
For the evaluation of such space-time integrals, one must take into account the
global geometry of our universe.
In this paper, we shall analyze the interplay of the global geometry and
integrals over quantum mechanical wave functions.

Clearly, considering this problem in full generality goes beyond the scope of
a single paper, and thus we must restrict attention to a more specific setting.
First of all, we only consider particles of
{\em{spin $\frac{1}{2}$}}, although the methods and results could easily be
extended to scalar fields or to particles of higher spin.
Furthermore, we focus on the physically relevant situation of the
{\em{Friedmann-Robertson-Walker}} (FRW) geometry, a spatially homogeneous and
isotropic space-time being the standard model for the present universe.
In the open and flat cases the spatial volume is infinite, so that the normalization integrals
of the wave functions will in general diverge. This is similar to the situation in Minkowski
space, where for the proper normalization one must first confine the particle to finite
$3$-volume, for example a finite box, and can then take the infinite volume limit
(see for example~\cite[\S2.6]{PFP}). This construction may depend on the form of the
finite box, whereas the global geometry of space-time does not seem to be relevant.
For these reasons, we shall not enter such constructions here.
The {\em{closed case}} is more interesting, because in this case the spatial volume
and thus also the normalization integrals are finite.
In other words, the global geometry gives rise to an {\em{infrared regularization}}.
To our knowledge, this effect and its consequences have not yet been studied.
The present paper aims at providing this analysis in reasonable generality and
sufficient detail.

We now introduce our problem more specifically and outline our results.
In so-called conformal coordinates, the line element of a spatially homogeneous and
isotropic space-time becomes
\begin{equation} \label{lineelement}
ds^2 = S(\tau)^2 \Big( d\tau^2 - d\chi^2 - f(\chi)^2(d\vartheta^2 + \sin^2{\vartheta} \:d\varphi^2) \Big) .
\end{equation}
Here $\tau$ is a time
coordinate, $\varphi \in [0, 2 \pi)$ and $\vartheta \in (0, \pi)$ are angular coordinates,
and~$\chi$ is a radial coordinate.
In the three cases of a closed, open and flat universe, the function~$f$
and the range of~$\chi$ are given respectively by \\[0.5em]
\centerline{ \begin{tabular}{lll}
closed universe: & $f(\chi) = \sin(\chi) \:,\quad$ & $\chi \in (0, \pi)$ \\
open universe: & $f(\chi) = \sinh(\chi) \:,\quad$ & $\chi >0 $ \\
flat universe:  & $f(\chi) = \chi \:,\quad$ & $\chi >0\:.$
\end{tabular} }

\vspace*{0.5em} \noindent
The function~$S$, the so-called {\em{scale function}}, is determined from the Einstein equations
and depends on the type of matter under consideration.
In the case of a perfect fluid, the function~$S$ must be a solution of the Friedmann equation,
and the resulting geometry is referred to as the FRW geometry.
In this paper, we do not need to specify the matter, and thus
the scale function can be an arbitrary positive function.

The Dirac operator on a Lorentzian manifold~$(M, g)$ reads
\begin{equation} \label{dop}
 {\mathcal{D}} = iG^j D_j\:,
 \end{equation}
where the Dirac matrices~$G^j$
are related to the metric by the anti-commutation relations
\begin{equation} \label{anticommute}
\{G^j,G^k\} \;\equiv\; G^j G^k + G^k G^j = 2 g^{jk} \1_{\C^4} \:,
\end{equation}
and~$D_j$ is the spin connection.
The Dirac equation takes the form
\begin{equation} \label{Dirac}
({\mathcal{D}} - m)\,\Psi = 0\:,
 \end{equation}
where~$\Psi$ is a four-component spinor, and~$m$ is the rest mass.
The spinors at every space-time point are endowed with an inner product~$\overline{\Psi} \Phi$
of signature~$(2,2)$ (where~$\overline{\Psi} =\Psi^\dagger \gamma^0$ is the usual adjoint spinor).
Integrating this inner product over space-time, we obtain the bilinear form
\begin{equation} \label{stip}
\boxed{ \quad \bra \Psi|\Phi \ket := \int_M \overline{\Psi} \Phi \: d\mu_M\:, \quad }
\end{equation}
where~$d\mu_M = \sqrt{|g|} d^4x$ is the volume element on~$M$
(here~$g=\det g_{ij}$ is the determinant of the metric).
Furthermore, choosing a space-like hypersurface~${\mathcal{H}}$, the probability integral
is given by
\begin{equation} \label{print}
\boxed{ \quad (\Psi|\Phi) := \int_{\mathcal{H}} \overline{\Psi} G^j \nu_j \Phi \:d\mu_{\mathcal{H}}\:,
\quad }
\end{equation}
where~$\nu$ is the future-directed normal on~${\mathcal{H}}$, and~$d\mu_{\mathcal{H}}$
is the volume element on~${\mathcal{H}}$.

The main purpose of this paper is to analyze both normalization integrals~\eqref{stip} and~\eqref{print}
for solutions of the Dirac equation in the closed universe, and to study the infinite volume
limit~$S \rightarrow \infty$. More precisely, after a brief introduction to the Dirac equation
in the FRW geometry and its separation (Section~\ref{sec2}), in Section~\ref{sec3}
we first compute the spatial normalization integral~\eqref{print}. Then, using a WKB-type
approximation in the time dependence, we can also make sense of the space-time integral~\eqref{stip}
(Proposition~\ref{prp2}). The validity of the WKB approximation is discussed at the end
of Section~\ref{sec3}.
These results allow us to introduce the so-called fermionic projector
as a well-defined projection operator on the generalized negative-energy solutions of the Dirac equation
(see~Section~\ref{sec4}, \eqref{Pproj}).
In Section~\ref{sec5} we consider the local form of the fermionic projector
in the infinite-volume limit and recover the Fourier integral describing the Dirac sea in the
Minkowski vacuum (see Theorem~\ref{thm2}).
Finally, in Appendix~\ref{appA} the spectrum and the eigenfunctions of the Dirac operator on~$S^3$
are computed in detail.

\section{The Dirac Equation in the FRW Geometry and its Separation} \label{sec2}
In this section we derive the Dirac equation in the FRW geometry and separate variables,
leaving us with an ODE describing the time dependence.
The Dirac equation in curved space-time was first formulated by Schr\"odinger~\cite{schroedinger}.
The more systematic study goes back to Brill and Wheeler~\cite{brill+wheeler}.
Today, there are different approaches, most notably the formalisms using null
frames~\cite{penrose+rindler} or pseudo-orthonormal frames~\cite{baum}.
The different formalisms yield different formulas for the Dirac operator,
which can be related to each other by suitable local transformations of the spinors
(see~\cite{schlueter} for a discussion of this point).
Here we use the approach in~\cite{U22}, which is most convenient for concrete calculations.
For an alternative derivation of the Dirac equation in the FRW geometry we refer to~\cite{Villalba}.
In order to satisfy the anti-commutation relations~(\ref{anticommute}), we choose the
Dirac matrices as
\beq \left. \begin{split}
 G^{\tau} &:= \frac{1}{S(\tau)}\: \gamma^0 \\
 G^{\chi} &:= \frac{1}{S(\tau)} \left( \cos\vartheta \ \gamma^3 +\sin\vartheta \cos\varphi \ \gamma^1 +
 \sin\vartheta \sin\varphi \ \gamma^2 \right) \\
G^\vartheta &:=  \frac{1}{S(\tau)\, f(\chi)} \left( -\sin\vartheta \ \gamma^3 +
 \cos\vartheta \cos\varphi \ \gamma^1 + \cos\vartheta \sin\varphi \ \gamma^2 \right) \\
G^\varphi &:= \frac1{S(\tau)\,f(\chi)\,\sin\vartheta} \left(-\sin\varphi \ \gamma^1 +
 \cos\varphi \ \gamma^2 \right) ,
\end{split}  \qquad \right\} \label{eq10}
 \eeq
where~$\gamma^0,\ldots, \gamma^3$ are the usual Dirac matrices of Minkowski space in
the Dirac representation,
\[  \gamma^0 = \left( \!\! \begin{array}{cc} \1 & 0 \\ 0 & -\1 \end{array} \!\! \right) , \qquad
\gamma^\alpha = \left( \!\!
            \begin{array}{cc}
            0 & \sigma^\alpha \cr
            -\sigma^\alpha & 0
            \end{array} \!\!
            \right) \:, \]
and~$\sigma^\alpha$, $\alpha=1,2,3$,  are the Pauli matrices,
\beq \label{PauliR3}
\sigma^1 = \begin{pmatrix} 0 & 1 \\ 1 & 0 \end{pmatrix}\:,\quad
\sigma^2 = \begin{pmatrix} 0 & -i \\ i & 0 \end{pmatrix}\:,\quad
\sigma^3 = \begin{pmatrix} 1 & 0 \\ 0 & -1 \end{pmatrix} .
\eeq
The spin connection~$D_j$ can be written as
\beq \label{Ej}
D_j = \partial_j -iE_j  \:,
\eeq
where~$E_j$, the so-called spin coefficients, are given in terms of the Dirac matrices
and their first derivatives. Substituting~(\ref{Ej}) into~(\ref{dop}), one sees that
in order to obtain the Dirac operator, it suffices to compute the combination~$G^j E_j$.
In the FRW geometry, this combination simplifies considerably:
\begin{Lemma} \label{lemma1}
For the choice of the Dirac matrices~(\ref{eq10}),
the spin coefficients satisfy the relation
\begin{equation} \label{GjEj}
G^jE_j = \frac i2 \:\nabla_j G^j\:,
\end{equation}
where~$\nabla$ denotes the Levi-Civita-connection.
\end{Lemma}
\Proof According to~\cite{U22}, the spin coefficients can be written as
\beq \label{Ejexplicit}
E_j = \frac i2 \:\rho (\partial_j \rho) -  \frac {i}{16} \:{\mbox{Tr}}(G^m \nabla_j G^n)G_m G_n   +    \frac i8 \:{\mbox{Tr}}(\rho G_j \nabla_m G^m)\, \rho\:,
\eeq
where
\[ \rho := \frac {i}{4!}\:\sqrt{|g|} \:\epsilon_{ijkl}\:G^iG^jG^kG^l \:, \]
and~$\epsilon_{ijkl}$ is the totally antisymmetric symbol
with~$\epsilon_{\tau \chi \vartheta \varphi}=1$.

Let us show that the first summand in~(\ref{Ejexplicit}) vanishes. Since the FRW metric is diagonal,
different Dirac matrices anti-commute, i.e.\
$G^iG^j=-G^jG^i$ if $i \neq j$. Using furthermore the anti-symmetry of the~$\epsilon$-symbol,
we conclude that
\begin{equation}
\rho = i \sqrt{|g|} \:G^\tau G^{\chi}G^{\vartheta}G^{\varphi}\:.
\end{equation}
Substituting the formula~$\sqrt{|g|} = S^4(\tau)f^2(\chi)\sin\vartheta$
and using the definitions~(\ref{eq10}), a straightforward calculation
shows that $\rho = i \gamma^0 \gamma^1 \gamma^2 \gamma^3$. Hence~$\rho$ is a constant,
and so its partial derivative in the first term in~(\ref{Ejexplicit}) vanishes.

Next we show that the last summand in~(\ref{Ejexplicit}) vanishes. Since the~$G^j$ are linear
combinations of the~$\gamma^j$, i.e.\ $G^n = {a^n}_j \gamma^j$ with real coefficients~$a^n_j$,
we find
\beq \label{nabG}
\nabla_n G^n = \nabla_n ({a^n}_j) \:\gamma^j\: ,
\eeq
and thus
\begin{displaymath}
{\mbox{Tr}}(\rho G_j \nabla_m G^m) = a^{i}_j \nabla_m ({a^m}_k) \:
{\mbox{Tr}}(\rho\, \gamma_i \gamma^k)\:.
\end{displaymath}
Due to the anti-commutation relations, the last trace vanishes for any choice of Dirac
matrices~$\gamma^i$ and~$\gamma^k$, proving that the last summand in~(\ref{Ejexplicit}) indeed
vanishes.

Substituting the remaining second summand in~(\ref{Ejexplicit}) into~(\ref{GjEj}), we obtain
\beq \label{Gj2}
G^j E_j = -\frac {i}{16} \:{\mbox{Tr}}(G_m \nabla_j G_n)\: G^j G^m G^n \:,
\eeq
where we raised and lowered the indices with the metric.
In the case where the indices~$m$, $j$, and~$n$ are all different, the expression is totally
anti-symmetric in these indices, because the last three Dirac matrices in~(\ref{Gj2}) anti-commute.
Consequently, in this case the symmetries of the Christoffel symbols
allow us to replace the covariant derivative by a partial derivative, and an explicit computation
of the partial derivatives shows that the trace vanishes. Hence it remains to consider the
case that at least two of the indices~$m$, $j$, and~$n$ coincide.
Ricci's lemma and the Leibniz rule yield that
\[ 0 = 4 \nabla_j g^{mn} = \nabla_j {\mbox{Tr}}(G^m G^n) =
{\mbox{Tr}}(\nabla_j G^m G^n) + {\mbox{Tr}}(G^m \nabla_j G^n) \:, \]
and thus we may anti-symmetrize the expression~(\ref{Gj2}) in the indices~$j$ and~$m$.
This gives rise to the simplification
\[ G^j E_j =  \frac i8 \:{\mbox{Tr}} (G^n \,\nabla_m G^m) \:G_n\:. \]
According to~(\ref{nabG}), the expression~$\nabla_m G^m$ can be written as a linear combination
of the Dirac matrices~$\gamma^j$, or alternatively of the matrices~$G^j$.
Applying the relation
\[ {\mbox{Tr}}(G^n G^j ) G_n = 4 g^{nj}\, G_n = 4 G^j \]
gives the result.
\QED

A major advantage of the relation~(\ref{GjEj}) is that it involves a divergence, which can be
computed using the Koszul formula
\[ \nabla_m G^m = \frac 1{\sqrt{|g|}}
\partial_m \Big(\sqrt{|g|} \,G^m \Big)\:, \]
making it unnecessary to compute the spin connection coefficients or even the
Christoffel symbols. A short calculation
combining Lemma~\ref{lemma1} with equations~(\ref{Ej}) and~(\ref{dop}) yields for the
Dirac equation in the FRW geometry
\beq \label{Dop}
\left[ iG^\tau \Big(\partial_\tau + \frac 32 \frac {\dot S}S \Big) + iG^\chi
\Big(\partial_\chi + \frac {f^\prime - 1}f \Big) +iG^\vartheta
\partial_\vartheta + i G^\varphi \partial_\varphi - m \right] \Psi = 0\:,
\eeq
where the dot and prime denote the partial derivatives with respect to~$\tau$ and~$\chi$,
respectively.
Multiplying by~$S(\tau)$, this equation can be written more conveniently as
\begin{equation}\label{DiracFRW}
\left[ i\gamma^0 \Big( \partial_\tau + \frac 32 \frac {\dot S}S \Big) - S(\tau) \ m + \left( \begin{array}{cc} 0 & {\mathcal{D}}_{\mathcal{H}} \\ -{\mathcal{D}}_{\mathcal{H}} & 0 \end{array} \right)   \right]
\Psi = 0 \:,
\end{equation}
where the purely spatial operator~${\mathcal{D}}_{\mathcal{H}}$ is given by
\begin{equation} \label{intrinsic}
{\mathcal{D}}_{\mathcal{H}} = i\sigma^\chi \left(
\partial_\chi + \frac {f^\prime - 1}f \right) +i\sigma^\vartheta
\partial_\vartheta + i \sigma^\varphi \partial_\varphi \:,
\end{equation}
and the matrices~$\sigma^\alpha$, $\alpha \in \{\chi, \vartheta, \varphi\}$, are linear combinations of the
Pauli matrices,
\beq \label{Pauli}
\left. \begin{split}
 \sigma^{\chi} &:= \cos\vartheta \ \sigma^3 +\sin\vartheta \cos\varphi \ \sigma^1 +
 \sin\vartheta \sin\varphi \ \sigma^2 \\
\sigma^\vartheta &:=  \frac{1}{f(\chi)} \left( -\sin\vartheta \ \sigma^3 +
 \cos\vartheta \cos\varphi \ \sigma^1 + \cos\vartheta \sin\varphi \ \sigma^2 \right) \\
\sigma^\varphi &:= \frac1{f(\chi)\,\sin\vartheta} \left(-\sin\varphi \ \sigma^1 +
 \cos\varphi \ \sigma^2 \right) .
 \end{split} \qquad \right\}
\eeq

In the case of a closed, open and flat universe, the operator~${\mathcal{D}}_{\mathcal{H}}$
is the Dirac operator on the sphere~$S^3$, on an hyperboloid, and on flat~$\R^3$,
respectively.
From now on, we shall restrict attention to the {\em{closed case}}, which has the advantage that
the spectrum of~${\mathcal{D}}_{\mathcal{H}}$ is discrete.
As is worked out in detail in Appendix~\ref{appA}, the operator~${\mathcal{D}}_{\mathcal{H}}$
can be identified with the intrinsic Dirac operator on~$S^3$. It is an essentially self-adjoint
 elliptic operator on the Hilbert space~$L^2(S^3)^2$ with domain
 of definition~$C^\infty(S^3)^2$ (see~\cite{LaMi}). It has the purely discrete spectrum~\cite{LaMi}
\beq \label{DS3spec}
\sigma({\mathcal{D}}_{\mathcal{H}}) = \left\{ \pm \frac{3}{2}, \,\pm \frac{5}{2}, \,\pm \frac{7}{2},
\ldots \right\} ,
\eeq
 and the dimension of the corresponding eigenspaces is
 \[ \dim \ker ( {\mathcal{D}}_{\mathcal{H}} - \lambda) = \lambda^2 - \frac{1}{4}\:. \]
In Appendix~\ref{appA} an orthonormal eigenvector basis is given explicitly
in terms of spherical harmonics and Jacobi polynomials (see Theorem~\ref{thmApp}).
It is denoted by~$(\psi^\pm_{njk})$,
where $n \in \N_0$, $j \in \N_0+\frac{1}{2}$ and ~$k \in \{-j, -j+1, \ldots, j\}$.
The eigenvalues are given by
\beq \label{evb}
{\mathcal{D}}_{\mathcal{H}} \psi^\pm_{njk} = \lambda\, \psi^\pm_{njk}\qquad \text{with} \qquad
\lambda = \pm(n+j+1)\:.
\eeq
More generally, we denote a normalized eigenfunction of~${\mathcal{D}}_{\mathcal{H}}$
corresponding to the eigenvalue~$\lambda$ by~$\psi_\lambda \in L^2(S^3)^2$.

Employing for~$\Psi$ the separation ansatz
\beq \label{sepansatz}
\Psi(\tau, \chi, \vartheta, \varphi) =
\frac1{S(\tau)^\frac32} \left( \!\! \begin{array}{c} h_1(\tau)\, \psi_\lambda(\chi, \vartheta, \varphi) \\
h_2(\tau)\, \tilde\psi_\lambda(\chi, \vartheta, \varphi) \end{array} \!\! \right) ,
\eeq
we obtain a coupled system of ODEs for the complex-valued functions~$h_1$ and~$h_2$,
\beq \label{timeODE}
\left[ i \partial_\tau - S m \left( \!\! \begin{array}{cc} 1 & 0 \\ 0 & -1 \end{array} \!\! \right)
+ \lambda \left( \!\! \begin{array}{cc} 0 & 1 \\ 1 & 0 \end{array} \!\! \right) \right]
\left( \!\! \begin{array}{c} h_1 \\ h_2 \end{array} \!\! \right) = 0 \:.
\eeq

\section{Normalization Integrals in the Closed FRW Geometry} \label{sec3}
In this section we shall analyze the integrals~\eqref{stip} and~\eqref{print}
for our separated wave function~(\ref{sepansatz}).
To compute the probability integral~\eqref{print}, we choose~$\mathcal{H}$ to be a slice of
constant conformal time $\tau$. Then the future-directed normal~$\nu$ has the components
$\nu_\tau = S(\tau)$ and $\nu_\alpha = 0$ for all $\alpha \in \{\chi, \vartheta, \varphi \}$.
The volume element on~$\mathcal{H}$ is
$d\mu_{\mathcal{H}} = \sqrt{|g_{\mathcal{H}}|} \ d\chi \
d\vartheta \ d\varphi = S^3(\tau) d\mu_{S^3}$, where~$d\mu_{S^3}$ is the
volume element on the unit sphere~$S^3$
(thus~$g_{\mathcal{H}}$ is the determinant of the induced metric). Hence
\beq \label{prob2}
(\Psi \,|\, \Psi) = \int_{\mathcal{H}} \Psi^\dagger \Psi\: d\mu_{\mathcal{H}}
= \int_{S^3} \left( |h_1|^2 + |h_2|^2 \right) |\psi_\lambda|^2\: d\mu_{S^3}
= |h_1|^2 + |h_2|^2\:.
\eeq
From the ODE~(\ref{timeODE}) one sees (using that the matrices in~(\ref{timeODE}) are all
Hermitian) that the function~$|h_1|^2 + |h_2|^2$ is constant in time. This corresponds to the more general
fact that, as a consequence of current conservation, the probability integral is independent of the choice
of $\mathcal{H}$.
Adopting the same convention as in~\cite[\S2.6]{PFP}, we normalize the wave functions such that
\beq \label{probnorm}
\boxed{ \quad (\Psi \,|\, \Psi) = \frac{1}{2 \pi}\:. \quad }
\eeq

The volume element in the space-time integral~\eqref{stip} is given by
\[ d\mu_M = \sqrt{|g|} \:d\tau \ d\chi \ d\vartheta \ d\varphi
= S d\tau \:d\mu_{\mathcal{H}}\:. \]
Thus for two wave functions~$\Psi$ and~$\tilde{\Psi}$,
\begin{align*}
\bra \Psi \,|\, \tilde{\Psi} \ket &= \int_M \overline{\Psi} \tilde{\Psi}\: d\mu_M
= \int S d\tau \int_{\mathcal{H}} d\mu_{\mathcal{H}}\: \overline{\Psi} \tilde{\Psi} \\
&= \int S d\tau \int_{\mathcal{H}} d\mu_{\mathcal{H}} \left( \overline{h_1} \tilde{h}_1
- \overline{h_2} \tilde{h}_2 \right) \: \langle \psi_\lambda, \psi_{\tilde{\lambda}}\rangle_{\C^2} \:,
\end{align*}
where the bar denotes complex conjugation. Hence
\beq \label{stip2}
\bra \Psi \,|\, \tilde{\Psi} \ket = \langle \psi_\lambda, \psi_{\tilde{\lambda}}\rangle_{L^2(S^3)^2}
\int \left( \overline{h_1} \tilde{h}_1 - \overline{h_2} \tilde{h}_2 \right) S d\tau \:.
\eeq

We first note that if the universe has a finite life time, then this normalization integral is
necessarily finite.
\begin{Prp} \label{prp31}
Consider a spatially homogeneous and isotropic geometry~\eqref{lineelement}
for the conformal time in the range~$\tau_{\min} < \tau < \tau_{\max}$
and singularities at~$\tau_{\min}$ (the ``big bang'') and at~$\tau_{\max}$ (the ``big crunch'').
Normalizing the wave functions according to~\eqref{probnorm}, the space-time inner
product~\eqref{stip} is finite and bounded by
\[ | \bra \Psi \,|\, \Psi \ket | \leq \frac{1}{2\pi} \int_{\tau_{\min}}^{\tau_{\max}} S\: d\tau\:. \]
\end{Prp}
\Proof Working in~\eqref{stip2} with normalized spatial eigenfunctions, we obtain
\[ \bra \Psi \,|\, \Psi \ket \leq  \int_{\tau_{\min}}^{\tau_{\max}} \left| |h_1|^2 - |h_2|^2 \right| S d\tau
\;\leq\;  \int_{\tau_{\min}}^{\tau_{\max}} \left( |h_1|^2 + |h_2|^2 \right) S d\tau\:. \]
The result follows immediately from~\eqref{prob2} and the normalization convention~\eqref{probnorm}.
\QED

Clearly, the time integral in~\eqref{stip2} is very large and will in general diverge in the limit
when the life time of the universe tends to infinity. In~\cite[\S2.6]{PFP} this problem is bypassed
by working with a variable mass parameter. In order to get the connection to this normalization method,
for a fixed spatial eigenvector~$\psi_\lambda$
we now consider solutions~$\Psi^m$ and~$\Psi^{m'}$ of the form~\eqref{sepansatz} for
two variable mass parameters~$m, m'>0$.
Then the corresponding time-dependent functions~$(h_1^m, h_2^m)$ and~$(h_1^{m'}, h_2^{m'})$
are solutions of~\eqref{DiracFRW}. We want to compute the inner product~$\bra \Psi^m | \Psi^{m'} \ket$, which according to~\eqref{stip2} becomes
\beq \label{stip3}
\bra \Psi^m | \Psi^{m'} \ket = \int_{\tau_{\min}}^{\tau_{\max}}
\left( \overline{h_1^m} h_1^{m'} -  \overline{h_2^m} h_2^{m'} \right) S\,d\tau\:.
\eeq
Since the Compton wave length is much smaller than the life time of our universe, the integrand 
will typically be highly oscillatory. Therefore, it seems appropriate to use a
{\em{WKB-type approximation}} (this approximation and its limitations will be discussed
in detail at the end of this section). To this end, we diagonalize the matrix potential
in~\eqref{timeODE} with a unitary matrix~$U(\tau)$,
\beq \label{diagonal}
U \begin{pmatrix} Sm & -\lambda \\ -\lambda & -Sm \end{pmatrix} U^{-1} =
\sqrt{m^2 S^2 +\lambda^2} \:\begin{pmatrix} 1 & 0 \\ 0 & -1 \end{pmatrix} .
\eeq
Then~\eqref{timeODE} can be written as
\beq \label{errorterm}
i \partial_t \left[ U \begin{pmatrix} h_1 \\ h_2 \end{pmatrix} \right] =
\sqrt{m^2 S^2 +\lambda^2} \,\begin{pmatrix} 1 & 0 \\ 0 & -1 \end{pmatrix} U 
\begin{pmatrix} h_1 \\ h_2 \end{pmatrix} + i \dot{U} \begin{pmatrix} h_1 \\ h_2 \end{pmatrix} .
\eeq
Under realistic conditions, the term~$\dot{U}$ is very small compared
to~$m S$. Leaving out this term, we can solve the ODE explicitly by
\beq \label{WKB}
\begin{pmatrix} h_1 \\ h_2 \end{pmatrix}\!(\tau) = U(\tau)^{-1}
\begin{pmatrix} \displaystyle c_1 \exp \left(-i \int^\tau \sqrt{m^2 S^2 +\lambda^2} \:d\tau \right) \\
\displaystyle  c_2 \exp \left(i \int^\tau \sqrt{m^2 S^2 +\lambda^2} \:d\tau \right) \end{pmatrix} .
\eeq
This approximation has the nice property that it respects current conservation
$|h_1|^2+|h_2|^2=\text{const}$, and thus the normalization~\eqref{probnorm}
is implemented simply by the condition
\beq \label{cnorm}
|c_1|^2 + |c_2|^2 = \frac{1}{2 \pi}\:.
\eeq
Indeed, this property is why we prefer~\eqref{WKB}
over the alternative method of rewriting the Dirac equation in terms of scalar second order equations
and employing the standard WKB ansatz.

Substituting the WKB ansatz~\eqref{WKB} into~\eqref{stip3} and multiplying out, we obtain
integrals of the form
\beq \label{oscillate}
\int_{\tau_{\min}}^{\tau_{\max}} \eta(\tau)
\exp \left\{ i \int_{\tau_0}^\tau \left(\mp \sqrt{m^2 S^2 +\lambda^2} \pm \sqrt{{m'}^2 S^2 +\lambda^2} \right) d\tilde{\tau} \right\} d\tau\:,
\eeq
where~$\tau_0$ determines the phase of the exponential, and~$\eta$ stands for the integration
density~$S$ times a combination of the
constants~$c^m_{1\!/\!2}$, $c^{m'}_{1\!/\!2}$ and certain matrix elements of the unitary
transformations~$U^m$ and~$U^{m'}$.
Qualitatively speaking, the functions~$\eta(\tau)$ and~$S(\tau)$ vary only on the cosmological scale,
whereas the exponential oscillates on the microscopic scale, unless the curly brackets in~\eqref{oscillate}
vanish. As a consequence of these rapid oscillations, the value of the integral will be small,
except if the square roots terms in~\eqref{oscillate} have opposite signs and~$m \approx m'$.
In order to quantify the contribution in this limit, in~\eqref{oscillate} we only take the
linear term in~$\delta m:=m-m'$ to obtain
\begin{align*}
I &:=  \int_{\tau_{\min}}^{\tau_{\max}} \eta(\tau) \exp \left\{ i \int_{\tau_0}^\tau
\left(\sqrt{m^2 S^2 +\lambda^2} - \sqrt{{m'}^2 S^2 +\lambda^2} \right) d\tilde{\tau} \right\} d\tau \\
&\, =  \int_{\tau_{\min}}^{\tau_{\max}} \eta(\tau)
\exp \left\{i \,\delta m \int_{\tau_0}^\tau \frac{m S^2}{\sqrt{m^2 S^2+\lambda^2}}\:
d\tilde{\tau} + {\mathscr{O}}((\delta m)^2)
 \right\} d\tau
 \end{align*}
(and similarly for the other combinations of signs).
After multiplying by a convergence generating factor~$e^{-\varepsilon (\delta m)^2}$, we can
integrate (for any fixed~$\tau$) over~$\delta m$ using the formula
\[ \int_\R e^{i \,Q\, \delta m - \varepsilon \, (\delta m)^2}\: d(\delta m) = \sqrt{\frac{\pi}{\varepsilon}}\:
e^{-\frac{Q^2}{4 \varepsilon}} \qquad \text{where} \qquad
Q = \int_{\tau_0}^\tau \frac{m S^2}{\sqrt{m^2 S^2+\lambda^2}}\:d\tilde{\tau} \:. \]
We thus obtain to leading order in~$\delta m$,
\beq \label{I2}
\int_\R I\: e^{-\varepsilon (\delta m)^2}\: d(\delta m) =
\sqrt{\frac{\pi}{\varepsilon}} \int_{\tau_{\min}}^{\tau_{\max}} \eta(\tau)
\exp \left( -\frac{Q(\tau)^2}{4 \varepsilon} \right) d\tau\: .
\eeq
In the limit~$\varepsilon \searrow 0$, the exponential tends to zero unless~$Q$ vanishes,
and thus we may evaluate the integral in the {\em{saddle point approximation}}.
Since the only zero of~$Q$ is at~$\tau=\tau_0$, we may replace~$Q$ in~\eqref{I2} by
\[ Q \approx Q'(\tau_0)\: (\tau-\tau_0) \:,\qquad
Q'(\tau_0) = \frac{m S(\tau_0)^2}{\sqrt{m^2 S(\tau_0)^2+\lambda^2}} \]
to obtain
\[ \int_\R I\: e^{-\varepsilon (\delta m)^2}\: d(\delta m) =
\sqrt{\frac{\pi}{\varepsilon}} \int_{\tau_{\min}}^{\tau_{\max}} \eta(\tau_0)\:
e^{-\frac{Q'(\tau_0)^2}{4 \varepsilon}\: (\tau-\tau_0)^2} d\tau
\xrightarrow{\varepsilon \searrow 0} 2 \pi \:\frac{\eta(\tau_0)}{Q'(\tau_0)}\:. \]
We conclude that in the above saddle point approximation, the oscillatory integral~\eqref{oscillate}
can be computed by
\begin{align}
\int_{\tau_{\min}}^{\tau_{\max}} & \eta(\tau)
\exp \left\{ i \int_{\tau_0}^\tau \left(s \sqrt{m^2 S^2 +\lambda^2} -s' \sqrt{{m'}^2 S^2 +\lambda^2} \right) d\tilde{\tau} \right\} d\tau \nonumber \\
&= 2 \pi \:\delta_{s,s'}\:\delta(m-m') \:\eta(\tau_0)
\:\frac{\sqrt{m^2 S(\tau_0)^2+\lambda^2}}{m S(\tau_0)^2}\:,
\label{resosc}
\end{align}
where~$s,s' \in \{1,-1\}$.
We remark that this result could be obtained more directly if one replaced the functions~$\eta$ and~$S$ in~\eqref{oscillate} by the constants~$\eta(\tau_0)$ and~$S(\tau_0)$, replaced the integral from~$\tau_{\min}$
to~$\tau_{\max}$ by an integral over the whole real line and applied the well-known
distributional formula
\[ \int_\R e^{i \omega t} = 2 \pi\:\delta(\omega)\:. \]
However, our method clarifies that the phase of the exponential in~\eqref{oscillate} as described
by the parameter~$\tau_0$ determines at what time the functions~$\eta$ and~$S$ in~\eqref{resosc}
are to be evaluated.

Evaluating the integral~\eqref{stip3} with the help of~\eqref{resosc} gives the following
result.
\begin{Prp} \label{prp2}
For a family of wave functions~$(\Psi^m)_{m>0}$ of the form~\eqref{sepansatz}
with a WKB ansatz for the time dependence~\eqref{WKB},
the inner product~\eqref{stip} can be computed in the saddle point approximation to be
\beq \label{normrel}
\bra \Psi^m \,|\, \Psi^{m'} \ket = 2 \pi \,\delta(m-m') \left( |c_1|^2 - |c_2|^2 \right) .
\eeq
\end{Prp}
\Proof Substituting~\eqref{WKB} into~\eqref{stip3}, according to~\eqref{resosc}
the mixed terms~$\sim \overline{c_1^m} c_2^{m'}$ and~$\sim \overline{c_2^m} c_1^{m'}$
do not contribute. Thus
\begin{align*}
\bra \Psi^m \,|\, \Psi^{m'} \ket = \int_{\tau_{\min}}^{\tau_{\max}}
&\Big\{ \overline{c_1^m} c_1^{m'} \left(\overline{(U^m)^1_1} (U^{m'})^1_1
-\overline{(U^m)^2_1} (U^{m'})^2_1 \right) e^{-i \Phi} \\
&\quad+ \overline{c_2^m} c_2^{m'} \left(\overline{(U^m)^1_2} (U^{m'})^1_2
-\overline{(U^m)^2_2} (U^{m'})^2_2 \right) e^{i \Phi} \Big\} \:S\, d\tau,
\end{align*}
where we used the abbreviation
\[ \Phi = \int_{\tau_0}^\tau \left( \sqrt{m^2 S^2 +\lambda^2} - \sqrt{m'^2 S^2 +\lambda^2} \right) 
d\tilde{\tau} \:. \]
Again using~\eqref{resosc}, we only get a contribution at~$m=m'$,
\begin{align}
\bra \Psi^m \,|\, \Psi^{m'} \ket \:=\: &
2 \pi \delta(m-m')\:  \frac{\sqrt{m^2 S^2+\lambda^2}}{m S} \nonumber \\
&\times \Big( |c_1|^2 \left( |U^1_1|^2 - |U^2_1|^2 \right)
- |c_2|^2 \left( |U^2_2|^2 - |U^1_2|^2 \right) \Big) , \label{Sdrop}
\end{align}
where we omitted the arguments~$\tau_0$ and the indices~$m$.
The unitary transformation~$U$ in~\eqref{diagonal} can be given explictly by
\[ U = \cos \!\left( \frac{\tau}{2} \right) \1 + \sin \!\left( \frac{\tau}{2} \right)
\begin{pmatrix} 0 & -1 \\ 1 & 0 \end{pmatrix} \qquad \text{with} \qquad
\tau = \arctan \!\left( \frac{\lambda}{mS} \right) . \]
Thus
\[ |U^1_1|^2 - |U^2_1|^2  = |U^2_2|^2 - |U^1_2|^2 = 
\cos^2 \!\left( \frac{\tau}{2} \right) - \sin^2 \!\left( \frac{\tau}{2} \right) =
\cos \tau =  \frac{mS}{\sqrt{m^2 S^2+\lambda^2}} \:, \]
giving the result.
\QED

We conclude this section by explaining the result of Proposition~\ref{prp2} and discussing its
physical significance.
Considering a variable mass parameter together with the $\delta(m-m')$-normalization~\eqref{normrel}
is very useful for the construction of wave functions for which the space-time integral~\eqref{stip}
is finite. More precisely, choosing~$\eta \in C^\infty_0((m-\varepsilon, m+\varepsilon))$, we can
``smear out'' the mass parameter on the scale~$\varepsilon$ by setting
\beq \label{msmear}
\Psi = \int \eta(m)\: \Psi^m \: dm
\eeq
(more generally, one can apply this method to composite expressions, see for example~\eqref{compsmear}). Then
\[ \bra \Psi \,|\, \Psi \ket = \int dm\, \overline{\eta(m)} \int dm'\, \eta(m') \:\bra \Psi^m \,|\, \Psi^{m'} \ket
= \int |\eta(m)|^2\: dm\:. \]

The surprising point about Proposition~\ref{prp2} is that the scale function~$S(\tau)$ drops out
when computing~\eqref{Sdrop}, making~\eqref{normrel} independent of the space-time geometry.
In particular, the parameter~$\tau_0$ describing the mass dependence of the phase in~\eqref{oscillate}
does not enter the $\delta(m-m')$-normalization. In this way, the result of Proposition~\ref{prp2}
is robust to the geometry of our homogeneous isotropic space-time and to the phases of the
function~$\eta(m)$ in the mass smearing~\eqref{msmear}.

Physically, the mass smearing~\eqref{msmear} can be understood as a technical device for introducing
a finite length scale~$\sim \varepsilon^{-1}$ for the correlation of~$\Psi$ with itself.
Such a finite correlation length could also arise as a consequence of an interaction or
self-interaction of a fully quantized system. In view of this more general picture, the conclusion
of the above construction is that if the correlation length is made finite in a Lorentz invariant way,
then the space-time integral~\eqref{stip} should be controlled explicitly in terms of the
probability integral~\eqref{print}.

Finally, the WKB approximation~\eqref{WKB} requires a detailed explanation.
To avoid misunderstandings, we first point out that we used the WKB ansatz only in the
time dependence, whereas the spatial dependence was treated without any approximations.
Consequently, the WKB approximation only enters the analysis of the time integral
in the space-time inner product~\eqref{stip}, giving the result of Proposition~\ref{prp2}.
The probability integral~\eqref{print}, however, was computed exactly (see~\eqref{probnorm}).
The time independence of the probability integral is a general consequence of current
conservation, without referring to any approximations.

Qualitatively speaking, the WKB wave function is a good approximation if the
potential is almost constant and non-zero (see for example~\cite{froman}).
More precisely, computing the error term in~\eqref{errorterm}, one finds that
the ansatz~\eqref{WKB} is a good approximation provided that the following condition holds:
\beq \label{WKBcond}
\frac{|\lambda|\, m}{(m^2 S^2 + \lambda^2)^{\frac{3}{2}}} \: \dot{S}\ll 1\:.
\eeq
For our present universe, the expression on the left is extremely small. Namely,
we can estimate it by
\[ \frac{|\lambda|\, m}{(m^2 S^2 + \lambda^2)^{\frac{3}{2}}} \: \dot{S} \leq
\frac{\dot{S}}{m S^2} \approx 2 \cdot 10^{-38}\:, \]
where we used that the quantity~$\dot{S}/S^2$ is the Hubble constant~$H \approx 2,3 \cdot 10^{-18}\,
\text{s}^{-1}$, whereas for~$m$ we set one over the Compton wave length of the electron,
$m \approx 1.24 \cdot 10^{20} \,\text{s}^{-1}$.
We conclude that at present, \eqref{WKB} is an extremely good approximation describing fermions
which move freely in the FRW geometry. Clearly, the WKB ansatz does not take into
account particular quantum effects like pair creation
(because the components of positive and negative frequency are not ``mixed'' in the
WKB ansatz; for a more detailed discussion of this point we refer to~\cite{fulling}).
But since we here focus on the gravitational interaction, such quantum effects
will be very small and should not be of relevance for all normalization issues.
Furthermore, the WKB approximation does not allow for the description of particular
quantum effects in the gravitational field like the Unruh or Hawking
radiation (see for example~\cite{wald}). However, these effects do not appear in the FRW geometry
and can thus be disregarded.

Another shortcoming of the WKB ansatz~\eqref{WKB} is that it does not apply
near the big bang or big crunch singularities. Indeed, as shown in~\cite{friedmann},
near these space-time singularities the Dirac spinors show quantum oscillations,
which cannot be described semi-classically and lead to
unexpected quantum effects which can even prevent the formation of the singularity.
Even in such rather wild quantum scenarios, the WKB approximation is of use and allows
us to compute the space-time integral~\eqref{stip}, as we now explain.
Proposition~\ref{prp31} (which does not use any approximations) tells us that the
space-time integral is a-priori bounded by~$\int S d\tau$. Thus we know that 
the time intervals shortly after the big crunch and shortly before the big bang
yield only very small contributions to the inner product~\eqref{stip} (no matter how wild the
behavior of the Dirac spinors and the metric is).
Away from the singularities, we can again use the WKB approximation to obtain the
result of Proposition~\ref{prp2}.

We conclude that, although the WKB approximation does certainly not account for all
physical effects of Dirac spinors in curved space-time,
for the purpose of analyzing the normalization integrals on the large scale,
the WKB ansatz~\eqref{WKB} seems an admissible and indeed very good approximation.

\section{The Global Normalization of the Fermionic Projector} \label{sec4}
We now turn attention to the normalization of the Dirac sea in the FRW geometry.
In the vacuum, the Dirac sea is composed of all negative-energy solutions of the Dirac equation.
In our time-dependent setting, the energy of the Dirac particles is not preserved in time, and thus
it is not obvious of which states the Dirac sea should be built up.
In~\cite{sea, grotz} this problem is resolved for a general time-dependent external field
by a global construction in space-time, giving a canonical splitting of the solution space into two
subspaces, one of which generalizes the negative-energy solutions of the vacuum and thus extends
the concept of the Dirac sea to the interacting situation.
However, because of technical assumptions on the decay of the external field at
infinity, these results do not immediately apply in the FRW metric, and the
constructions of~\cite{sea, grotz} have not yet been extended to the FRW geometry.
Fortunately, the problem of distinguishing the ``generalized negative energy solutions''
disappears in the WKB approximation~\eqref{WKB}, because then the
negative-energy part is obtained simply by setting~$c_1=0$.
Working again with the approximation~\eqref{WKB}, we can
describe the Dirac sea explicitly, as we now explain.

Following~\cite{sea, grotz}, the {\em{fermionic projector}}~$P$ is an operator
on the Dirac wave functions whose image is spanned by the states of the Dirac sea. Furthermore, the
fermionic projector should be symmetric and idempotent with respect to the inner product~\eqref{stip}.
Thus, using a bra/ket-notation,
\beq \label{Pdef}
P = -\sum_{\lambda, \sigma} |\Psi_{\lambda, \sigma} \ket \bra \Psi_{\lambda, \sigma}| \:,
\eeq
where~$\lambda$ refers to the eigenvalues of the spatial Dirac operator~$\D_{S^3}$, and
$\sigma$ labels a basis of the corresponding eigenspaces of~$\D_{S^3}$,
and the Dirac wave functions must satisfy the normalization condition
\[ \bra \Psi_{\lambda', \sigma'} |\Psi_{\lambda, \sigma} \ket = \delta_{\lambda, \lambda'}
\delta_{\sigma, \sigma'}\:. \]
It is convenient to write~\eqref{Pdef} with an integral kernel,
\[ (P \Psi)(x) = \int_M P(x,y)\: \Psi(y)\: d\mu_M \quad \text{with} \quad
P(x,y) = -\sum_{\lambda, \sigma} \Psi_{\lambda, \sigma}(x) \:\overline{\Psi_{\lambda, \sigma}(y)}\:. \]
Employing the ansatz~\eqref{sepansatz} and working for the time dependence with the WKB
approximation~\eqref{WKB}, we select the generalized negative-energy solutions by
choosing~$c_1=0$. In order to satisfy~\eqref{cnorm}, we choose~$c_2=(2 \pi)^{-\frac{1}{2}}$.
Clarifying the dependence on the mass parameter by an index~$m$,
we thus obtain for the kernel of the fermionic projector
\begin{align}
P&_m(x,y) = - \frac{1}{2 \pi} \left( S(x^0)\: S(y^0) \right)^{-\frac{3}{2}}
\sum_{\lambda, \sigma} \psi_{\lambda, \sigma}(\vec{x})\:
\overline{\psi_{\lambda, \sigma}(\vec{y})} \nonumber \\
&\otimes \left[ U(x^0)^{-1} \begin{pmatrix} 0 & 0 \\ 0 & 1 \end{pmatrix} U(y^0)
\begin{pmatrix} 1 & 0 \\ 0 & -1 \end{pmatrix} \right]
\exp \left(i \int_{x^0}^{y^0} \sqrt{m^2 S^2 +\lambda^2} \:d\tau \right) , \label{Pproj}
\end{align}
where the~$\psi_{\lambda, \sigma}$ form an orthonormal basis of the $\lambda$-eigenspace
of~$\D_{S^3}$ and $x^0$, $y^0$ denote the time coordinates.

Clearly, the above operator~$P_m$ is symmetric with respect to the inner product~\eqref{stip}.
Furthermore, using the normalization results~\eqref{cnorm} and~\eqref{normrel}, it is
idempotent if we work with a $\delta$-normalization,
\[ P_m \,P_{m'} = \delta(m-m')\: P_m\:. \]
This $\delta$-normalization can be avoided similar to~\eqref{msmear} by ``smearing out''
the mass parameter, for example by setting for any~$\delta>0$
\[ P = \int_{m-\delta}^{m+\delta} P_{\mu} d\mu \:. \]
Then this~$P$ really is idempotent,
\beq \label{compsmear}
P^2 = \int_{m-\delta}^{m+\delta} d\mu
\int_{m-\delta}^{m+\delta} d\mu'\; P_m P_{m'}
=  \int_{m-\delta}^{m+\delta} P_m = P \:.
\eeq

\section{The Local Form of the Fermionic Projector} \label{sec5}
We shall now explore how the fermionic projector~\eqref{Pproj} looks like locally, in
the physically realistic case when the size of the universe is much larger than the
length scale for local observations. Our analysis involves the two approximations
that the scale function~$S$ is very large and that it is almost constant locally.

\begin{Thm} \label{thm2}
We consider the fermionic projector~\eqref{Pdef} with the individual states
normalized according to~\eqref{probnorm}.
Then in the limit~$S \rightarrow \infty$ of a large universe and neglecting local curvature effects,
the fermionic projector~\eqref{Pdef} in a local
reference frame takes the standard form of Minkowski space,
\beq \label{Pmres}
P_m(x,y) = \int \frac{d^4k}{(2 \pi)^4} \: (k \slsh + m)\: \delta(k^2-m^2)\:\Theta(-k^0)\: e^{-ik(x-y)}\:.
\eeq
\end{Thm}
\Proof For convenience, we choose the local reference frame such that~$y=0$.
In order to bring the Fourier integral in~\eqref{Pmres} into a more convenient form,
we set~$\omega=k^0$, $t=x^0$, $r=|\vec{x}|$ and choose polar coordinates~$(p=|\vec{k}|,
\vartheta,\varphi)$ in momentum space with the north pole pointing in the direction of~$\vec{x}$
(thus~$\vartheta$ denotes the angle between~$\vec{k}$ and~$\vec{x}$).
We then obtain
\begin{align}
&\int \frac{d^4k}{(2 \pi)^4}\: (k \slsh + m)\: \delta(k^2-m^2)\:\Theta(-k^0)\: e^{-ikx} \nonumber \\
&= \frac{1}{(2 \pi)^4} \:(i \Pdd_x + m) \int_{-\infty}^\infty \!\!d\omega
\int_0^\infty \!\!p^2\, dp \int_{-1}^1 \!\! d\cos \vartheta
\int_0^{2 \pi} \!\!d\varphi\: \delta(\omega^2 - p^2-m^2)\: e^{-i \omega t+i p r \cos \vartheta} \nonumber \\
&= \frac{1}{(2 \pi)^3} \:(i \Pdd_x + m) \int_0^\infty \frac{p^2 dp}{2 |\omega(p)|}\:e^{-i \omega(p) t}\:
\int_{-1}^1 e^{i p r \cos \vartheta}\: d\cos \vartheta \nonumber \\
&= \frac{1}{(2 \pi)^3} \:(i \Pdd_x + m) \int_0^\infty \frac{p\, dp}{|\omega(p)|}
\:\frac{\sin(pr)}{r}\:e^{-i \omega(p) t} \nonumber \\
&= \frac{1}{(2 \pi)^3} \int_0^\infty dp\:\frac{p}{|\omega(p)|}
\left\{ ( \omega(p) \gamma^0 + m )\: \frac{\sin(pr)}{r}
+ i \gamma^r \partial_r \left( \frac{\sin(pr)}{r}\right) \right\} e^{-i \omega(p) t}\:, \label{goal}
\end{align}
where we set~$\omega(p)=-\sqrt{p^2+m^2}$ and
\beq \label{grdef}
\gamma^r = \cos\vartheta \ \gamma^3 +\sin\vartheta \cos\varphi \ \gamma^1 +
\sin\vartheta \sin\varphi \ \gamma^2 \:.
\eeq

Our goal is reproduce~\eqref{goal} from~\eqref{Pproj} by suitable approximations.
First, neglecting local curvature effects, we may assume that~$S$ is constant in
our neighborhood. Furthermore, we may disregard the
$\tau$-dependence of the transformation~$U$. From~\eqref{diagonal} one sees that
\[ U^{-1} \begin{pmatrix} 0 & 0 \\ 0 & 1 \end{pmatrix} U
= \frac{\1}{2} - \frac{1}{\sqrt{m^2 S^2 + \lambda^2}} \begin{pmatrix} S m & -\lambda \\
-\lambda & -Sm \end{pmatrix} . \]
Moreover, rewriting the spatial eigenfunctions in~\eqref{Pproj}
in terms of the spectral projectors~$E_\lambda$
as computed in the appendix (see~\eqref{Eldef} and~\eqref{Ekern}), we obtain
\begin{align}
&P_m(x,y) \nonumber \\
&\;= \frac{1}{2 \pi S^3} \sum_{\lambda} \frac{1}{2\, |\omega|}\:E_\lambda(\vec{x},\vec{y})
 \otimes \left[ \omega \begin{pmatrix} 1 & 0 \\ 0 & -1 \end{pmatrix}
+ \frac{\lambda}{S} \begin{pmatrix} 0 & 1 \\ -1 & 0 \end{pmatrix} + m \right]
e^{i \omega S (y^0-x^0)} \:,  \label{Pmrep}
\end{align}
where
\[ \omega = -\sqrt{\frac{\lambda^2}{S^2} + m^2} \:. \]
In order to further simplify this formula, we fix~$y$ and choose our coordinate system
such that~$y^0=0$ and that the point~$\vec{y} \in S^3$ is at the north pole.
Describing the point~$x$ by the coordinates~$(\tau, \chi, \vartheta, \varphi)$, the spatial
spectral projectors can be expressed by Lemma~\ref{lemmaA6}.
Next, it is important to observe that the quantities~$\tau$, $\chi$ and~$\lambda$ are
dimensionless. In order to obtain the usual physical observables time, radius and momentum, we
must multiply by suitable powers of~$S$,
\[ t = S \tau\:, \qquad r = S \sin \chi \approx S \chi \:,\qquad p = \frac{|\lambda|}{S}\:. \]
These local observables should be well-defined in the limit~$S \rightarrow \infty$.
This means in particular that the spatial eigenvalues~$|\lambda|$ must tend to infinity.
This allows us to simplify the formula of Lemma~\ref{lemmaA6}
using the large-$n$-asymptotics of the Jacobi polynomials~\cite[eq.~22.15.1]{AS} and
the Stirling formula~\cite[eq.~6.1.37]{AS}. This gives
\[ E_{\pm(n+\frac{3}{2})}(x, y)
= \frac{1}{4 \pi^2} \left[ \frac{n \sin(n \chi)}{\chi} \pm i \sigma^\chi \left( \frac{n \cos(n \chi)}{\chi}
- \frac{\sin(n \chi)}{\chi^2} \right) \right] \left(1 + {\mathscr{O}} \!\left( \frac{1}{n} \right)\right) . \]
Substituting this formula in~\eqref{Pmrep}, and multiplying out, we can rewrite the
tensor products of $2 \times 2$-matrices in terms of Dirac matrices
(also compare~\eqref{Pauli} and~\eqref{grdef}),
\[ \1_{\C^2} \otimes \begin{pmatrix} 1 & 0 \\ 0 & -1 \end{pmatrix} = \gamma^0\:, \qquad
\sigma^\chi \otimes \begin{pmatrix} 0 & 1 \\ -1 & 0 \end{pmatrix} = \gamma^r\:. \]
Moreover, inspecting the signs in the case of positive and negative~$\lambda$,
one sees that both cases give the same contributions.
Thus the fermionic projector becomes
\[ P_m(x,y) = \frac{1}{8 \pi^3 S^3} \sum_{n \in \N_0} \frac{n}{|\omega|}
\left[ (\omega \gamma^0 + m) \frac{\sin(n \chi)}{\chi} +
i p \gamma^r \left( \frac{\cos(n \chi)}{\chi}
- \frac{\sin(n \chi)}{n \chi^2} \right) \right]
e^{-i \omega t} \:, \]
up to corrections of higher order in~$1/n$.
Using that~$|\lambda|=n+\frac{3}{2}$, we can set~$p=n/S$. Furthermore,
considering the $n$-series as a Riemann sum, one sees that the sum
goes over to an integral via
\beq \label{riemann}
\frac{1}{S} \sum_{n \in \N_0} \cdots \xrightarrow{S \rightarrow \infty}
\int_0^\infty dp  \cdots\:,
\eeq
and at the same time the corrections of higher order in~$1/n$ tend to zero.
We thus recover~\eqref{goal}.
\QED

We point out that the same normalization constant was obtained in~\cite[\S2.6]{PFP} by considering
the system in finite $3$-volume in Minkowski space and taking the infinite-volume limit.
We here obtain the same result using a more realistic infrared regularization
in the form of the closed FRW geometry. Our result shows that despite the naive
divergence of the space-time integral~\eqref{stip}, the fermionic projector is well-defined
in the infinite volume limit. Our analysis also confirms that the normalization
constant is independent of the details of the regularization procedure.

We remark that the above methods could also be used to analyze in detail how the
fermionic projector in the FRW geometry deviates
from that of Minkowski space. Namely, the local curvature is captured
by the Jacobi polynomials. The effect of the global geometry could be analyzed by considering
the difference of the integral and the Riemann sum in~\eqref{riemann}. However, this analysis would
go beyond the scope of this paper.

\section{Conclusion}
We saw that the Dirac equation in the closed FRW geometry can be separated into ODEs describing
the spatial and the time dependence. The spatial dependence can be solved
in closed form (see Appendix~\ref{appA}). This makes it possible to compute
the probability integral~\eqref{print}, which is time independent
due to current conservation.
For the space-time normalization integral~\eqref{stip} the situation is more difficult,
because the solution of the time-dependent ODE cannot be given in closed form.
Thus we must rely on estimates (Lemma~\ref{prp31}) as well as on
the WKB approximation~\eqref{WKB}, which are justified at the end of Section~\ref{sec3}.
After ``smearing out'' the mass parameter (see~\eqref{msmear}), we can make sense of the
space-time integral~\eqref{stip} (Proposition~\ref{prp2}).
We apply these results to construct the so-called fermionic projector, being
a well-defined projection operator on the generalized negative-energy solutions of the Dirac equation
(see~\eqref{Pproj}). Locally, the fermionic projector has the same form
as in Minkowski space (see Theorem~\ref{thm2}).
In this way, a proper normalization of the fermionic states was obtained
in a physically realistic setting. We also learn that the details of the global geometry do affect neither
the local form of the fermionic projector nor the relation between the normalization integrals (\ref{stip}) and (\ref{print}).

\appendix
\section{Spectrum and Eigenfunctions of the Dirac Operator on~$S^3$} \label{appA}
In this appendix we derive the intrinsic Dirac operator on~$S^3$ (with the standard metric),
compute its eigenvalues and construct an explicit orthonormal eigenvector basis in terms of
special functions. We also derive convenient formulas for the spectral projectors.
Clearly, the results of this section are not new.
But it seems worth to give a self-contained and explicit treatment, in particular
because the detailed formulas for the spectral projectors are needed in Section~\ref{sec5}.
As in~\eqref{lineelement}
we choose coordinates~$(\chi, \vartheta, \varphi) \in (0, \pi) \times (0, \pi) \times (0, 2 \pi)$, such that
the line element on~$S^3$ becomes
\[ ds^2 = d\chi^2 +\sin^2\chi \ d\vartheta^2 + \sin^2\chi \ \sin^2\vartheta \ d\varphi^2 \:. \]
We first verify that the operator~(\ref{intrinsic}) really is the intrinsic Dirac operator.
\begin{Lemma} The intrinsic Dirac operator on~$S^3$ can be written as
\begin{equation} \label{DS3}
{\mathcal{D}}_{S^3} \;=\; i\sigma^\chi\left(\partial_\chi + \frac {\cos \chi - 1} {\sin \chi}\right) +i\sigma^\vartheta
\partial_\vartheta + i \sigma^\varphi \partial_\varphi \:,
\end{equation}
where the matrices~$\sigma^\chi$, $\sigma^\vartheta$ and~$\sigma^\varphi$ are given by~(\ref{Pauli}).
\end{Lemma}
\Proof Before we can apply the methods of~\cite{U22}, we must get back to a Lorentzian manifold.
To this end, we consider the manifold~$M=\R \times S^3$ with the line element~$d\tilde{s}^2$ given by
\beq \label{leprod}
d\tilde{s}^2 = d\tau^2 - ds^2 \:,
\eeq
where~$\tau \in \R$ is the ``time coordinate'' and~$ds^2$ again denotes the metric on~$S^3$.
Then the hypersurface~$\tau=0$ is isometric to~$S^3$, and the corresponding second fundamental
form vanishes identically. As a consequence, the Dirac operator on~$M$ can be written as
a constant time derivative term plus a term involving the intrinsic Dirac operator on~$S^3$,
\beq \label{Dtriv}
{\mathcal{D}}_M = i \begin{pmatrix} \1 & 0 \\ 0 & -\1 \end{pmatrix} \partial_t
+ \left( \begin{array}{cc} 0 & {\mathcal{D}}_{S^3} \\ -{\mathcal{D}}_{S^3} & 0 \end{array} \right).
\eeq
The line element~\eqref{leprod} is obtained from that of the closed FRW metric~\eqref{lineelement}
by setting~$S \equiv 1$. In this special case, the Dirac operator in~\eqref{Dop}
reduces to the Dirac operator in~\eqref{DiracFRW} with~$S \equiv 1$. Comparing with~\eqref{Dtriv}
gives the result.
\QED

Our first step in diagonalizing the operator~\eqref{DS3} is to separate the~$\vartheta$- and~$\varphi$-dependence in~\eqref{DS3}.
The next lemma gives a connection to the standard angular momentum operators
\beq \label{LKdef}
\vec{L} =  -i \vec{x} \wedge \vec{\nabla} \qquad \text{and} \qquad
K = \vec{\sigma} \vec{L} + 1\:,
\eeq
where~``$\wedge$'' denotes the cross product in~$\R^3$.
\begin{Lemma} The angular Dirac matrices in~\eqref{Pauli} and the angular momentum
operators~\eqref{LKdef} satisfy the relations
\begin{align}
\sigma^\vartheta \partial_\vartheta + \sigma^\varphi \partial_\varphi &= -
\frac{\sigma^\chi}{\sin \chi} \:\vec{\sigma} \vec{L} \label{1rel} \\
K \sigma^\chi &= - \sigma^\chi K\:. \label{2rel}
\end{align}
\end{Lemma}
\Proof Choosing in Euclidean~$\R^3$ the polar coordinates~$(\sin\chi, \vartheta, \varphi)$
with~$\sin\chi=|\vec{x}|$ , a short calculation shows that

\begin{align*}
\sigma^\vartheta \partial_\vartheta + \sigma^\varphi
\partial_\varphi &= \vec{\sigma} \vec{\nabla} - \sigma^\chi \partial_{\sin\chi}
= \frac{\sigma^\chi}{\sin \chi}\;\vec{\sigma}\vec{x}
\left(\vec{\sigma} \vec{\nabla} - \sigma^\chi \partial_{\sin\chi} \right) \\
(\vec{\sigma} \vec{x})(\vec{\sigma} \vec{\nabla}) &= \vec{x} \vec{\nabla}
+ i \vec{\sigma} (\vec{x} \wedge \vec{\nabla}) = \sin \chi \,\partial_{\sin\chi} - \vec{\sigma} \vec{L}\:.
\end{align*}
Combining these relations gives~\eqref{1rel}. Furthermore, using the Leibniz rule as well as
the anti-commutation relations of the Pauli matrices, we obtain
\[ \sin\chi \left( \vec{\sigma} \vec{L} \,\sigma^\chi+ \sigma^\chi\, \vec{\sigma} \vec{L} \right)
= -i \left\{ \vec{\sigma} (\vec{x} \wedge \vec{\nabla}) , \:x^\alpha \sigma^\alpha \right\}
= -2 i \vec{x} (\vec{x} \wedge \vec{\nabla})
-i \vec{\sigma} (\vec{x} \wedge \vec{\sigma}) = -i \vec{\sigma} (\vec{x} \wedge \vec{\sigma})\:, \]
where we used the Einstein summation convention for the index~$\alpha$.
Writing the cross product with the totally antisymmetric Levi-civita
symbol~$\epsilon^{\alpha \beta \gamma}$ and using the relation
\[ \sigma^\alpha \sigma^\beta = i \epsilon^{\alpha \beta \gamma} \sigma^\gamma\:, \]
we find
\[ -i \vec{\sigma} (\vec{x} \wedge \vec{\sigma}) = -i \: \epsilon^{\alpha \beta \gamma}
\sigma^\gamma x^\alpha \sigma^\beta = x^\alpha
\epsilon^{\alpha \beta \gamma} \epsilon^{\gamma \beta \delta}
\sigma^\delta = -2 \vec{x} \vec{\sigma} \:, \]
and thus
\[ \vec{\sigma} \vec{L} \,\sigma^\chi+ \sigma^\chi\, \vec{\sigma} \vec{L} = \frac{-2 \vec{x} \vec{\sigma}}{\sin\chi}\:. \]
Using the definition of~$K$ in~\eqref{LKdef} we obtain~\eqref{2rel}.
\QED
Following the procedure in quantum mechanical textbooks, we next diagonalize the
operator~$K$ in terms of the spherical harmonics~$Y^k_l$.
\begin{Lemma} \label{lemmaA3}
The two-component wave functions~$\chi^k_{j \mp \frac 12} \in L^2(S^2)^2$
with~$j \in \N_0+\frac{1}{2}$ and~$k \in \{-j, -j+1,...,j\}$ given by
\beq
\begin{split}
\chi^k_{j-\frac 12} &= \sqrt{ \frac {j+k}{2j} } \: Y \normalsize ^{k- \frac 12}_{j-\frac 12}
\begin{pmatrix} 1 \\ 0 \end{pmatrix} + \sqrt{ \frac {j-k}{2j} } \: Y \normalsize ^{k + \frac 12}_{j-\frac 12} \begin{pmatrix} 0 \cr 1 \end{pmatrix} \\
\chi^k_{j+\frac 12} &= \sqrt{ \frac {j+1-k}{2j+2} } \: Y \normalsize ^{k- \frac 12}_{j+\frac 12} \begin{pmatrix} 1 \\ 0 \end{pmatrix} - \sqrt{ \frac {j+1+k}{2j+2} } \: Y \normalsize ^{k + \frac 12}_{j+\frac 12} \begin{pmatrix} 0 \\ 1 \end{pmatrix}
\end{split}  \label{angular}
\eeq
form an orthonormal eigenvector basis of the operator~$K$ in~\eqref{LKdef}, with the eigenvalues given by
\begin{equation} \label{Kchi}
K \chi^k_{j \mp \frac 12} = \pm \left(j + \frac 12 \right) \chi^k_{j\mp\frac 12}\:.
\end{equation}

Furthermore,
\beq \label{sigchi}
\sigma^\chi \chi^k_{j \mp \frac 12} = \chi^k_{j \pm \frac 12}\:.
\eeq
\end{Lemma}
\Proof Following standard conventions, the spherical harmonics~$Y^k_l(\vartheta,\varphi) \in L^2(S^2)$ with~$l \in \N_0$ and~$k \in \{-l, -l+1, \ldots, l\}$ are an eigenvector basis of the angular momentum operators~$L^2$
and~$L_z$,
\[ L^2 Y^k_l = l(l+1) Y^k_l \:,\qquad L_z Y^k_l = k Y^k_l \:. \]
As a consequence, the ``ladder operators''~$L_{\pm} := L_x \pm iL_y$ satisfy the relations
\[ L_{\pm} Y^k_l = \sqrt{l(l+1)-k(k \pm 1)} \:Y^{k \pm 1}_l \:, \]
where for convenience we used the convention~$Y^k_l = 0$ if~$|k| > l$.
Using these relations together with the explicit form of the Pauli matrices~\eqref{PauliR3},
a straightforward calculation gives~\eqref{Kchi}  (for more details see~\cite{sakuraiadv}).
The orthonormality of the wave functions~\eqref{angular} is verified directly using that the spherical harmonics are orthonormal in~$L^2(S^2)$. The completeness of the~$\chi^k_{j \pm \frac{1}{2}}$
follows because taking suitable linear combinations of the wave functions in~\eqref{angular}
we obtain the spinors
\[ Y^{k- \frac 12}_{j+\frac 12} \begin{pmatrix} 1 \\ 0 \end{pmatrix}
\qquad \text{and} \qquad
Y^{k + \frac 12}_{j-\frac 12} \begin{pmatrix} 0 \cr 1 \end{pmatrix} , \]
which clearly form a basis of~$L^2(S^2)^2$.

In order to explain~\eqref{sigchi}, we first apply~\eqref{2rel} to obtain
\[ K \sigma^\chi \chi^k_{j \mp \frac 12} = -\sigma^\chi K \chi^k_{j \mp \frac 12} = \mp
\left( j+\frac12 \right) \sigma^\chi \chi^k_{j \mp \frac 12} \:. \]
Hence the operator~$\sigma^\chi$ maps the eigenspaces of~$K$ corresponding to
the eigenvalues~$\pm (j+\frac{1}{2})$ into each other. Since this operator is unitary,
it is clear that the vectors~$(\sigma^\chi \chi^k_{j \mp \frac 12})_{k=-j,\ldots, j}$
form an orthonormal basis of the eigenspace corresponding to the eigenvalue~$\mp (j+\frac{1}{2})$.
This proves~\eqref{sigchi} up to unitary transformations of the eigenspaces.
To verify that these unitary transformations are the identity, one needs to go through
a straightforward computation using the detailed form of the wave functions~\eqref{angular}
and of the matrix~$\sigma^\chi$ as given in~\eqref{Pauli}.
\QED

We are now ready to separate the Dirac equation on~$S^3$. In the Dirac equation
\[ {\mathcal{D}}_{S^3} \psi = \lambda \psi\, \]
we substitute~\eqref{DS3} and multiply from the left by~$-i \sigma^\chi$.
Using~\eqref{1rel} and~\eqref{LKdef}, the Dirac equation becomes
\[ \left( \partial_\chi + \frac {\cos \chi - 1} {\sin \chi} - \frac{K-1}{\sin \chi}
+ i \lambda \sigma^\chi \right) \psi = 0\:. \]
Next, for any~$j \in \N_0+\frac{1}{2}$ and~$k \in\{-j, -j+1,\ldots, j\}$ we take the ansatz
\begin{equation} \label{sepansatz2}
\psi(\chi, \vartheta, \varphi) =  \frac{1}{\sin \chi} \left(
\Phi_1(\chi) \: \chi^k_{j - \frac12}(\vartheta, \varphi) -i
\Phi_2(\chi) \: \chi^k_{j + \frac12}(\vartheta, \varphi) \right)
\end{equation}
with two complex functions~$\Phi_1$ and~$\Phi_2$.
Applying~\eqref{Kchi} and~\eqref{sigchi}, we obtain the system of ODEs
for the two-spinor~$\Phi=(\Phi_1, \Phi_2)$,
\begin{equation} \label{radialODE1}
\left[ \partial_\chi - \frac {j+\frac12}{\sin \chi}\: \begin{pmatrix} 1 & 0 \\ 0 & -1 \end{pmatrix}
+ \lambda \left( \begin{array}{cc} 0 & 1 \cr -1 &
0 \end{array} \right) \right] \Phi = 0\:.
\end{equation}
We refer to this system as the {\em{radial equations}}. It can also be written as the eigenvalue problem
\begin{equation} \label{radialODE2}
{\mathcal{R}} \Phi = \lambda \Phi
\qquad \text{where} \qquad
{\mathcal{R}} = \begin{pmatrix} 0 & 1 \\ -1 & 0 \end{pmatrix} \partial_\chi + \frac {j+\frac12}{\sin \chi}\: \begin{pmatrix} 0 & 1 \\ 1 & 0 \end{pmatrix} .
\end{equation}
Using the orthonormality of the eigenfunctions~$\chi^k_{j \pm \frac{1}{2}}$, we find that
\beq \label{L2norm}
\langle \psi, \psi \rangle_{L^2(S^3)^2} = \int_0^\pi |\Phi|^2 \: d\chi\:.
\eeq
Hence our goal is to find all solutions to~\eqref{radialODE2} which are normalizable
with respect to the standard $L^2$ scalar product on the interval~$(0, \pi)$.

In the special case~$\lambda=0$, the ODEs~\eqref{radialODE1} can easily be solved in
closed form to obtain the two fundamental solutions
\[ \tan^{j+\frac{1}{2}} \left( \frac{\chi}{2} \right) \begin{pmatrix} 1 \\ 0 \end{pmatrix}
\qquad \text{and} \qquad
\cot^{j+\frac{1}{2}} \left( \frac{\chi}{2} \right) \begin{pmatrix} 0 \\ 1 \end{pmatrix} \]
Every non-trivial linear combination of these two function has a non-square-integrable singularity
at~$\chi=0$ or~$\chi=\pi$, and thus~$\lambda=0$ is not an eigenvalue.
In the remaining case~$\lambda \neq 0$, we can solve the second equation in~\eqref{radialODE1}
for~$\Phi_1$,
\beq \label{Phi1eq}
\Phi_1 = \frac{1}{\lambda} \left( \Phi_2' + \frac{j+\frac{1}{2}}{\sin \chi}\: \Phi_2 \right) .
\eeq
Differentiating this relation and substituting both~$\Phi_1$ and~$\Phi_1'$ in the first
equation in~\eqref{radialODE1}, we obtain the second order scalar equation
\beq
\Phi_2'' + \left( - \frac{j+\frac{1}{2}}{\sin^2 \chi} \left(\cos \chi +j+\frac{1}{2} \right)
 +\lambda^2 \right) \Phi_2 = 0\:. \label{Phi2}
\eeq

We next construct the general solution to the ODEs~\eqref{Phi2}.
We take the ansatz $\Phi_2(\chi) = h(\chi)\, g(\chi)$ with
\[ h(\chi) = \left( \frac{1+\cos\chi}{1-\cos\chi} \right)^{\frac {j + \frac12} 2} \]
and perform the coordinate transformation
\beq \label{trafo}
\chi \rightarrow y:=\frac{1-\cos\chi}{2}\:.
\eeq
This leads to the hypergeometric differential equation
 \[ (y^2-y)\, g^{\prime \prime}(y) + (y+(2j+1))\, g^\prime(y) - \lambda^2\, g(y) = 0 \:, \]
whose general solution can be written in terms of the hypergeometric function~${_2}F_1$ by
\beq \label{gdef}
g(y) = \alpha\: {_2}F_1(-\lambda,\lambda,-j,y) \:+\: \beta\: y^{j+1} \,{_2}F_1(-\lambda+j+1,\lambda+j+1,2+j,y)
\eeq
with two complex parameters~$\alpha$ and~$\beta$. We conclude that the general solution
to~\eqref{Phi2} is
\begin{equation}\label{radialsoln2}
\Phi_2(\chi)=\left( \frac{1+\cos\chi}{1-\cos\chi}\right)^{\frac{j+\frac12}{2}} g\left(\frac{1-\cos\chi}{2}\right) .
\end{equation}

We next evaluate the condition that~$\Phi_2$ must be square integrable.
Near~$\chi=0$ we can use the relation ${_2}F_1(a,b,c,0)=1$ to obtain the expansion
\[ \Phi_2 = \left( \alpha\:2^{j+\frac{1}{2}}\: \chi^{-j-\frac{1}{2}}
+  \beta\:2^{-j-\frac{3}{2}}\: \chi^{j+\frac{3}{2}} \right) (1+{\mathscr{O}}(\chi)) \:, \]
showing that the parameter~$\alpha$ must vanish.
We conclude that~$\Phi_2$ must be of the form
\beq \label{Phi22}
\Phi_2 =  \beta\: \left( 1-y \right)^{\frac{j}{2}+\frac{1}{4}}
\left( y \right)^{\frac{j}{2}+\frac{3}{4}}
\:{_2}F_1(-\lambda+j+1,\lambda+j+1,2+j,y) \:.
\eeq
To avoid trivialities, we can assume
that the remaining parameter~$\beta$ is non-zero (namely, otherwise~$\Phi_2$ would vanish
identically and, according to~\eqref{Phi1eq}, $\Phi_1$ would also be identically zero).
We then obtain a non-square-integrable pole at~$\chi=\pi$ unless the corresponding hypergeometric function~${_2}F_1$ in~\eqref{Phi22} vanishes at~$y=1$.
From the expansion~\cite[eq.~15.4.23]{DLMF}
one sees that this is the case only if
\[ \frac{\Gamma(2+j)}{\Gamma(j+1-\lambda) \:\Gamma(j+1+ \lambda)} = 0 \:. \]
Since the $\Gamma$-function has no zeros (see~\cite[eq.~6.1.12]{AS}),
this condition can only be satisfied if the denominator is singular.
Using that the Gamma function $\Gamma(x)$ is only singular at the negative
integers~$x=0,-1,-2,\ldots$ (see~\cite[after eq.~6.1.3]{AS}), we obtain the following necessary condition.
\begin{Lemma} \label{lemmaA4}
The radial equations~\eqref{radialODE1} admits a square-integrable solution
$\Phi \in L^2((0, \pi))^2$ only if~$\lambda$ is of the form
\beq \label{lambdadef}
\lambda=\pm (n+j+1) \qquad \text{with~$n \in \mathbb{N}_0$}\:.
\eeq
\end{Lemma} \noindent
The result of this lemma greatly simplifies our formulas, because for these admissible values of
~$\lambda$, the hypergeometric function in~\eqref{gdef} can be expressed in terms of a Jacobi polynomial
(see~\cite[eq.~15.4.6]{AS}),
\begin{align}
\Phi_2(\chi) &= \beta \left( \frac{1-\cos\chi}{2}\right)^{\frac{j}{2}+\frac{3}{4}}
\left( \frac{1+\cos\chi}{2}\right)^{\frac{j}{2}+\frac{1}{4}}\: \frac{n!}{(2+j)_n}\: P_n^{(j+1,j)}(\cos\chi)
\nonumber \\
&= \frac{\beta \,n!}{2^{j+1}\,(2+j)_n}\:
\left( 1-u\right)^{\frac{j}{2}+\frac{3}{4}} \left( 1+u\right)^{\frac{j}{2}+\frac{1}{4}}\: P_n^{(j+1,j)}(u)\:,
\label{phi2}
\end{align}
where~$(2+j)_{n}$ denotes the Pochhammer symbol, and in the last line we set~$u=\cos \chi$.
Substituting this result into~\eqref{Phi1eq} we can compute the derivative using~\cite[eq.~22.8.1]{AS}.
Furthermore, we can use~\cite[eqns.~22.7.17--22.7.19]{AS} to modify the prefactors~$(1\pm u)$
as well as the upper indices of the Jacobi polynomials. By suitably applying these relations,
one obtains
\beq \label{phi1}
\Phi_1(\chi) = \pm \frac{\beta \,n!}{2^{j+1}\,(2+j)_n}\:
\left( 1-u\right)^{\frac{j}{2}+\frac{1}{4}} \left( 1+u\right)^{\frac{j}{2}+\frac{3}{4}}\: P_n^{(j,j+1)}(u)\:,
\eeq
where the signs~$\pm$ refer to the two choices in~\eqref{lambdadef}.
The $L^2$-norm of the functions~\eqref{phi2} and~\eqref{phi1} can be computed
with the rule (see~\cite[eq.~22.2.1]{AS})
\[ \int_{-1}^1 (1-u)^a\: (1+u)^b\: \left(P_n^{(a,b)}(u)\right)^2 = \frac{2^{a+b+1}}{2n+a+b+1}\;
\frac{\Gamma(n+a+1) \,\Gamma(n+b+1)}{n!\, (n+a+b)!}\:. \]
In particular, one sees that~$\Phi_1$ and~$\Phi_2$ are both square integrable.
Choosing the constant~$\beta$ such that the $L^2$-norm of~$\Phi$ (as given by~\eqref{L2norm}) equals
one, we obtain
\beq
\begin{pmatrix} \Phi_1 \\ \Phi_2 \end{pmatrix}_{\pm,j,n} =
\frac{\sqrt{n!\, (n+2j+1)!}}{2^{j+\frac{1}{2}}\, \Gamma(j+2)}\: \sin^{j+\frac{1}{2}}(\chi)
\begin{pmatrix}
\cos(\chi/2)\: P_{n}^{(j, j+1)}(\cos \chi) \\[0.5em]
\pm \sin(\chi/2)\: P_{n}^{(j+1,j)}(\cos \chi) \end{pmatrix} . \label{Phiresult}
\eeq

We finally summarize our results.
\begin{Thm} \label{thmApp} The intrinsic Dirac operator on~$S^3$ has the
spectrum~\eqref{DS3spec}. Working in the representation~\eqref{DS3}, an orthonormal eigenvector
basis~$\psi^\pm_{njk}$ with~$n \in \N_0$, $j \in \N_0 + \frac{1}{2}$
and~$k \in \{-j, -j+1,\ldots, j\}$ is obtained by the ansatz~\eqref{sepansatz}
with~$\Phi = (\Phi_1, \Phi_2)$ according to~\eqref{Phiresult}.
The corresponding eigenvalues are given by~\eqref{evb}.
\end{Thm}
\Proof Noting that~${\mathcal{D}}_{S^3}$ is an elliptic differential operator
on the compact manifold~$S^3$, standard elliptic theory (see~\cite{LaMi}) yields
that~${\mathcal{D}}_{S^3}$ is essentially self-adjoint and has a purely discrete spectrum.
Since in Lemma~\ref{lemmaA3} we constructed a basis for the angular dependence, and in
Lemma~\ref{lemmaA4} we determined all eigenvalues
of the angular operator, the resulting eigenfunctions~$\psi^\pm_{njk}$ are clearly a basis
of~$L^2(S^3)^2$. Furthermore, according to Lemma~\ref{lemmaA3} and the normalization
after~\eqref{phi1}, the basis vectors all have norm one. Thus it remains to show that
the functions~$\psi^\pm_{njk}$ are orthogonal. This follows immediately from the
orthogonality of the angular functions~$\chi^k_{j\pm \frac{1}{2}}$ (see Lemma~\ref{lemmaA3})
as well as the fact that eigenvectors corresponding to different eigenvalues of~${\mathcal{D}}_{S^3}$
are orthogonal.
\QED

We finally derive a useful formula for the spectral projectors. For any eigenvalue~$\lambda$,
we denote the projector on the corresponding eigenspace by~$E_\lambda$. It can be
expressed as an integral operator
\beq \label{Eldef}
(E_\lambda \psi)(x) = \int_{S^3} E_\lambda(x,y) \: \psi(y) \:d\mu_y\:,
\eeq
where~$x,y \in S^3$ and~$d\mu$ is the volume form on~$S^3$.
In terms of the orthonormal basis~$\psi^\pm_{njk}$ of Theorem~\ref{thmApp}, the
kernels become
\beq \label{Ekern}
E_{\pm|\lambda|}(x,y) = \sum_{n=0}^{|\lambda|-\frac{3}{2}} \;\sum_{k=-j}^j \psi^\pm_{njk}(x)
\overline{\psi^\pm_{njk}(y)} \Big|_{j=|\lambda|-n-1}\:,
\eeq
where the bar denotes the adjoint spinor with respect
to the scalar product defined pointwise on the spinors.
This expression simplifies considerably if the second argument is evaluated at the north pole,
$y=\mathfrak{n}$.

\begin{Lemma} \label{lemmaA6}
The integral kernels of the spectral projectors~$E_\lambda$ in~\eqref{Eldef}
satisfy for any~$n \in \N_0$ the relations
\[ E_{\pm(n+\frac{3}{2})}(x, \mathfrak{n})
= \frac{(n+2)!}{8 \pi^{\frac{3}{2}}\, \Gamma(n+\frac{3}{2})}
\left( \cos(\chi/2)\:P^{(\frac{1}{2}, \frac{3}{2})}_n(\chi) \mp i \sigma^\chi \sin(\chi/2)\: P^{(\frac{3}{2},
\frac{1}{2})}_n(\chi)\right) , \]
where~$\mathfrak{n}$ denotes the north pole $\chi=0$, $x \in S^3$ is parametrized by the
coordinates~$(\chi, \vartheta, \varphi)$, and~$\sigma^\chi$ is defined in~\eqref{Pauli}.
\end{Lemma}
\Proof
In view of~\eqref{Phiresult} and~\eqref{sepansatz2}, $\psi^\pm_{njk}(\mathfrak{n})$ vanishes
unless~$j=\frac{1}{2}$. In this case, from~\eqref{Phiresult} and~\cite[eq.~22.2.1]{AS} we
conclude that
\[ \lim_{\chi \rightarrow 0} \frac{\Phi_1(\chi)}{\sin \chi}  = \sqrt{\frac{(n+2)!}{\pi\,n!}}\qquad \text{and} \qquad
\lim_{\chi \rightarrow 0} \frac{\Phi_2(\chi)}{\sin \chi}  = 0 \:. \]
Furthermore, in the case~$j=\frac{1}{2}$, the $2$-spinors~\eqref{angular} simplify to
\begin{align*}
\chi^{\frac{1}{2}}_{\frac{1}{2}-\frac{1}{2}} &= \frac{1}{\sqrt{4 \pi}}
\begin{pmatrix} 1 \\ 0 \end{pmatrix}, &
\chi^{-\frac{1}{2}}_{\frac{1}{2}-\frac{1}{2}} &= \frac{1}{\sqrt{4 \pi}}
\begin{pmatrix} 0 \\ 1 \end{pmatrix} \\
\chi^{\frac{1}{2}}_{\frac{1}{2}+\frac{1}{2}} &= \frac{1}{\sqrt{4 \pi}}
\begin{pmatrix} \cos \vartheta \\ e^{i \varphi} \sin \vartheta \end{pmatrix}, &
\chi^{-\frac{1}{2}}_{\frac{1}{2}+\frac{1}{2}} &= \frac{1}{\sqrt{4 \pi}}
\begin{pmatrix} e^{-i \varphi} \sin \vartheta \\ -\cos \vartheta \end{pmatrix}\:,
\end{align*}
and thus
\[ \sum_{k=\pm \frac{1}{2}}
\chi^k_{\frac{1}{2}-\frac{1}{2}}\, \overline{\chi^k_{\frac{1}{2}-\frac{1}{2}}}
= \frac{1}{4 \pi} \: \1 \:,\qquad
\sum_{k=\pm \frac{1}{2}}
\chi^k_{\frac{1}{2}+\frac{1}{2}}\, \overline{\chi^k_{\frac{1}{2}-\frac{1}{2}}}
= \frac{1}{4 \pi} \: \sigma^\chi \:. \]
Using these formulas in~\eqref{sepansatz2} and~\eqref{Ekern}, we obtain
\[ E_{\pm(n+\frac{3}{2})}(x,\mathfrak{n}) =
\frac{1}{4 \pi\, \sin \chi} \Big( \Phi_1(\chi) \:\1 - i \Phi_2(\chi) \:\sigma^\chi \Big) \Big|_{j=\frac{1}{2}}
\sqrt{\frac{(n+2)!}{\pi\,n!}}\:, \]
and substituting~\eqref{Phiresult} gives the result.
\QED

\Thanks{{\em{Acknowledgments:}}  We would like to thank A.\ Grotz, C.\ Morris
and the referees for their careful reading and helpful comments on the manuscript.}

\providecommand{\bysame}{\leavevmode\hbox to3em{\hrulefill}\thinspace}
\providecommand{\MR}{\relax\ifhmode\unskip\space\fi MR }
\providecommand{\MRhref}[2]{%
  \href{http://www.ams.org/mathscinet-getitem?mr=#1}{#2}
}
\providecommand{\href}[2]{#2}

\end{document}